\documentclass{article}

\usepackage[margin=3cm]{geometry}
\usepackage{amsmath,amssymb,amsfonts}
\usepackage[nocompress]{cite}
\usepackage{orcidlink}
\usepackage{hyperref}
\hypersetup{colorlinks=true,breaklinks=true,linkcolor=blue,filecolor=magenta,urlcolor=cyan}

\usepackage{authblk}
\newcommand{\keywords}[1]{\textbf{Keywords:} #1}

\begin{document}

\title{Buchdahl stars and bounds with cosmological constant}
\setlength{\affilsep}{3ex}

\author[1,3]{Christian G. B\"ohmer\footnote{Email: c.boehmer@ucl.ac.uk}\orcidlink{0000-0002-9066-5967}}
\author[2,3]{Naresh Dadhich\footnote{Naresh Dadhich passed away on 6 Nov 2025, in Beijing, China while on an academic visit. This paper was last discussed with him on 23 Sep 2025.}\orcidlink{0000-0002-4439-9071}}
\author[4,5]{Surajit Das\footnote{Email: surajitdas@mail.ustc.edu.cn}\orcidlink{0000-0003-2994-6951}}
\affil[1]{Department of Mathematics, University College London, \protect\\
Gower Street, London WC1E 6BT, United Kingdom\protect\vspace{1ex}}
\affil[2]{Inter University Centre for Astronomy \& Astrophysics, \protect\\
Post Bag 4, Pune 411007, India\protect\vspace{1ex}}
\affil[3]{Astrophysics Research Centre, School of Mathematics, Statistics \protect\\
and Computer Science, University of KwaZulu-Natal, \protect\\
Private Bag X54001, Durban 4000, South Africa\protect\vspace{1ex}}
\affil[4]{Department of Astronomy, School of Physical Sciences, \protect\\
University of Science and Technology of China, Hefei, Anhui 230026, China\protect\vspace{1ex}}
\affil[5]{CAS Key Laboratory for Researches in Galaxies and Cosmology, \protect\\
School of Astronomy and Space Science, \protect\\
University of Science and Technology of China, Hefei, Anhui 230026, China}

\date{22 May 2026} 

\maketitle

\begin{abstract}
The Schwarzschild interior solution, when combined with the assumption of a finite central pressure, leads to the well-known Buchdahl bound. This bound establishes an upper limit on the mass-to-radius ratio of an object, which is equivalent to imposing an upper limit on the gravitational potential. Remarkably, this limit exhibits considerable universality, as it applies to a broader class of solutions beyond the original Schwarzschild interior metric. By reversing this argument, one can define the most compact horizonless object that satisfies this gravitational bound. Intriguingly, the same bound arises when applying the Virial theorem to an appropriately chosen combination of gravitational and potential energy. In this work, we explore the generalised Buchdahl compactness bound in the presence of a cosmological constant. We investigate its implications, define a suitable gravitational energy and an associated potential energy that incorporate the cosmological term, and demonstrate that the universality of the Buchdahl bound persists. However, we also observe that different bounds emerge depending on the chosen approach.
\end{abstract}
\bigskip\bigskip

\clearpage

\section{Introduction}
\label{s1}

One of the fundamental questions in astrophysics concerns the maximum possible compactness achievable by a self-gravitating object within the framework of Einstein's general relativity (GR). This pivotal question was first explored by Buchdahl in 1959 by examining perfect fluid spheres~\cite{Buchdahl:1959zz,Buchdahl:1966star}. These findings demonstrated that for self-gravitating, spherically symmetric perfect fluid solutions in GR, the compactness obeys the upper limit $GM/(Rc^2)\leq 4/9 \approx 0.444$, where $M$ and $R$ denote the object's mass and radius, respectively. This bound was derived under fairly general assumptions, including non-negative density and isotropic pressure, while further assuming that the density is non-increasing ($d\rho/dr \leq 0$). This interior solution is then matched to the exterior Schwarzschild solution at the boundary surface $R_b$, which is defined by the vanishing pressure condition $p(R_b)=0$. The limit arises when one assumes the central pressure to satisfy $P_c < \infty$. Alternatively, the same limit was derived by imposing the strong energy condition ($p_r + 2p_t - \rho \leq 0$), where $p_r$, $p_t$, and $\rho$ denote radial pressure, transverse pressure, and matter energy density, respectively~\cite{Andreasson:2007ck,Karageorgis:2007cy}. Here, the bound is saturated for an infinitely thin shell with the condition $2p_t = \rho_m$. On the other hand, Chandrasekhar derived the theoretical upper limit for the mass and density of white dwarfs, which are approximately $M_{\rm max}\approx 1.44M_\odot$ and $\rho_{\rm max}\approx10^9~{\rm g/cm^3}$, respectively~\cite{Chandrasekhar:1931ih,Shapiro:1983du}. Consequently, the maximum mass plays a vital role in distinguishing between different relativistic astrophysical objects (such as neutron stars and black holes) within compact binary systems. Additionally, it is essential in determining the final states of various astrophysical phenomena, including supernova collapses and binary neutron star mergers. Although neither scenario is entirely physical, they serve as idealised limiting cases. 

In contrast, the event horizon of a Schwarzschild black hole is defined by $R=2GM/c^2$, corresponding to the compactness limit equal to $1/2$. Consequently, Buchdahl's theorem establishes a critical restriction: within GR, fluid configurations cannot attain compactness values arbitrarily close to the black hole limit. This result serves as a fundamental benchmark for investigating the nature of compact relativistic astrophysical objects~\cite{Cardoso:2019rvt}.

Various alternative approaches have investigated different scenarios to derive compactness bounds, including the study of charged static objects~\cite{Giuliani:2007zza,Mak:2001ie,Boehmer:2007gq} and their subsequent extensions~\cite{Andreasson:2009qu,Lemos:2015wfa,Lemos:2010te,Arbanil:2014usa}. Other studies have explored these limits in the context of brane-world gravity~\cite{Germani:2001du, Garcia-Aspeitia:2014pna}, relaxed versions of Buchdahl's initial assumptions~\cite{Karageorgis:2007cy,Andreasson:2008xw}, and the inclusion of a cosmological constant $\Lambda$~\cite{Mak:2000asc,Harko:2000ac,Andreasson:2012dj,Stuchlik:2000gey}. Additionally, investigations have been conducted in modified gravity frameworks~\cite{Goswami:2015dma}, including $f(R)$ gravity theories~\cite{Fernandez:2025cfp}, as well as Lovelock gravity~\cite{Dadhich:2016fku,Dadhich:2010qh} and higher-dimensional spacetimes~\cite{PoncedeLeon:2000pj,Zarro:2009gd,Wright:2015yda}. Recently, a generalised maximum mass limit was derived by considering the most general equation of state for elastic matter with constant longitudinal wave speeds~\cite{Alho:2022bki}. This work demonstrated that enforcing subluminal wave propagation gives the maximum compactness of any static elastic star to approximately $0.462$. Furthermore, it was shown that incorporating radial stability constraints lowers this bound to around $0.389$. Thus, the Buchdahl limit serves as a universal benchmark under broad physical conditions, though more compact configurations remain possible under certain circumstances.

This paper investigates the generalised Buchdahl compactness inequality in the context of a non-zero cosmological constant $\Lambda$. Among the diverse cosmological parameters, the cosmological constant holds a crucial significance in understanding the dynamics of the universe. Modern observational data strongly indicate the presence of a positive cosmological constant $\Lambda > 0$~\cite{Planck:2018vyg,DES:2021wwk,Riess:2021jrx,Vagnozzi:2021quy}. The so-called `concordance' models, which study cosmic microwave background fluctuations and Hubble constant measurements, favour a spatially flat universe, a prediction consistent with inflationary theory~\cite{Ostriker:1995su,Krauss:1995yb}. Comparative analyses reveal that models incorporating a positive $\Lambda$ are significantly preferred over those without it, as evidenced by studies involving globular cluster age estimates and baryon abundance constraints~\cite{Krauss:1997wy}. Furthermore, observations of high-redshift Type Ia supernovae indicate a negative deceleration parameter, offering direct empirical confirmation for a repulsive cosmological constant~\cite{SupernovaCosmologyProject:1996grv}. Alternatively, the cosmological constant can be interpreted as a measure of the intrinsic temperature of vacuum spacetime or as a representation of the thermal properties inherent to the geometry itself~\cite{Gasperini:1987fi,Gasperini:1988jq}.

We aim to explore the compactness limit solely based on the unique exterior solution for uncharged Schwarzschild-de Sitter objects. While compactness typically depends on the internal fluid properties and binding energy, the universality of gravity implies a possible connection between the interior and exterior properties of the object. In the exterior region, the gravitational field energy, which can be computed at any point outside the object, may play a pivotal role in this analysis. This energy represents the gravitational energy contained outside a compact object of radius $R$ and decreases monotonically to zero as $R \to \infty$. In the Newtonian approximation, it is given by $-M^2/(2R)$, analogous to the electrostatic energy $Q^2/(2R)$, where $Q$ denotes the charge of the object. Although this quantity is extensive and inherently negative, its absolute value can also be considered. Notably, the compactness ratio $GM/(Rc^2)$ for a compact object of mass $M$ and radius $R$, much like the gravitational field energy, decreases monotonically to zero at infinity. This analogy suggests that a limit on compactness could equivalently be formulated as a constraint on the gravitational field energy. Consequently, one of our co-authors recently proposed a compactness limit defined purely in terms of gravitational field energy, which is entirely determined by the exterior metric, without any dependence on the interior~\cite{Dadhich:2019jyf}. Furthermore, it is well-established that the Buchdahl compactness bound for any static relativistic astrophysical object requires the gravitational field energy to be no greater than half of its non-gravitational matter energy~\cite{Dadhich:2019jyf,Dadhich:2023csk}. More broadly, we propose that the Buchdahl compactness limit universally applies to any static relativistic astrophysical compact object—whether charged or neutral—provided its gravitational field energy does not exceed half of its non-gravitational matter energy, effectively enforcing the Virial theorem. Following Ref.~\cite{Dadhich:2019jyf}, we derive the Buchdahl-analogue compactness bound in the presence of a cosmological constant using the framework of gravitational field energy and the Virial theorem. This result precisely coincides with the Buchdahl bound obtained from the condition on the surface potential $\Phi(R) \leq 4/9$. The Buchdahl inequality of a perfect fluid implies that the limit $M/R\to 4/9$ cannot be reached as this would require $P_c\to\infty$. However, it was shown in~\cite{Andreasson:2006qp,Andreasson:2006ab} that Vlasov matter can saturate this bound. By assuming staticity and spherical symmetry, a sequence of static shells of the Einstein-Vlasov system with this property was shown to exist.

It is thus reasonable to consider the question: what can be said about the interior solution of a Schwarzschild-de Sitter object? To address this, we examine the Buchdahl star bound in the presence of a cosmological constant, as previously explored~\cite{Boehmer:2003uz,Balaguera-Antolinez:2004ytv,Boehmer:2005sm,Balaguera-Antolinez:2005ngl} and compare the bounds of the interior and the exterior solutions.

The paper is organised as follows: We begin with a brief review of the Misner-Sharp-Hernandez mass function and the Brown-York quasi-local energy for a static neutral relativistic astrophysical compact object described by the Schwarzschild-de Sitter metric in Sec.~\ref{s2}. From these definitions, in Sec.~\ref{s3} we derive the gravitational field energy. Subsequently, we establish the compactness limit based on the gravitational field energy prescription and the Virial theorem, followed by a comparison with the constant density interior solution. Finally, we conclude with a brief discussion. We work with units $G=c=1$.

\section{Basic definitions and preliminaries}
\label{s2}

Let us begin with a static and spherically symmetric line element in curved spacetime using Schwarzschild-like coordinates given by
\begin{eqnarray}
    ds^2 = -N^2(R) dt^2 + \frac{dR^2}{f^2(R)} + R^2 d\Omega^2\,,
    \label{1}
\end{eqnarray}
where $N(R)$ and $f(R)$ are known as the redshift function and the shape function, respectively, and $d\Omega^2 = d\theta^2 + \sin^2\negmedspace\theta\, d\phi^2$ is the line element of a unit 2-sphere. In the context of general relativity, the concept of energy (or, particularly gravitational field energy), is an ambiguous concept. Since gravitational energy is associated with the curvature of spacetime, it is fundamentally non-localisable, making it challenging to define with clarity and precision. However, in cases of high symmetry, such as static, spherical symmetry, certain physically plausible formulations exist~\cite{Brown:1992br,Liu:2003bx}.

\subsection{Misner-Sharp-Hernandez mass}
\label{s2.1}

The Misner-Sharp-Hernandez (MSH) mass is focused on the mass-energy of a gravitational system in a quasilocal manner, which is utilised in studies of the gravitational collapse of fluids~\cite{Misner:1964je,Hernandez:1966zia,Hayward:1994bu}. In the form of the metric~\eqref{1}, the MSH mass function is defined as follows:
\begin{eqnarray}
    M_{\rm MSH} := \frac{R}{2}\big(1-f^2\big)\,,
    \label{2}
\end{eqnarray}
where the definition of the MSH mass function follows from a gauge-independent geometrical scalar definition, defined locally in terms of the areal radius $R$ such that $(1-2M_{\rm MSH}/R)=\nabla_\mu R\nabla^\mu R$~\cite{Misner:1964je,Hernandez:1966zia,Stephani:2003tm}. It is crucial to mention that the MSH mass (or equivalently energy) of a perfect fluid reduces to the Newtonian mass in a leading order in the Newtonian limit. Geometrically it is a measure of an average value of potential energy~\cite{Dadhich:2023csk}. Moreover, for the case of the Schwarzschild solution, the Schwarzschild black hole mass $M$ matches with the mass $M_{\rm MSH}$ of that MSH solution, which suggests that a system with an infinite dispersion of bare mass $M$ starts to collapse due to the influence of its own gravitational pull. On the other hand for the Reisser-Nordstr\"om background, the effective gravitational mass (as referred here $M_{\rm MSH}$) of a charged object with total charge $Q$ at any radius $R$ is expressed as $(M-Q^2/(2R))$, where the electrostatic energy outside of radius $R$ has been deducted from bare mass $M$.

At this point, one should ask whether the mass definition~\eqref{2} is suitable in the presence of a cosmological constant. As discussed, for example in~\cite{Boehmer:2007ae}, it is equally well-motivated to define
\begin{align}
    \widetilde{M} = \frac{R}{2}(1-\nabla^\mu R \nabla_\mu R) - \frac{\Lambda}{6}R^3 = M_{\rm MSH} - \frac{\Lambda}{6}R^3 \,.
    \label{2a}
\end{align}
Here, one explicitly subtracts the contribution of the cosmological term. This ensures that the $\Lambda$ term drops out when calculating the mass in spacetimes which are asymptotically de Sitter or anti-de Sitter. The physical results are, of course, independent of the chosen mass definition, however, the different definitions need to be distinguished as their interpretations differ slightly.

\subsection{Brown-York mass}
\label{s2.2}

The prescription of the Brown-York (BY) quasilocal mass function, again based on~\eqref{1}, is given by~\cite{Brown:1992br}
\begin{align}
    M_{\rm BY} := R\big(1-f\big) \,.
    \label{3}
\end{align}
In general, the BY mass is formulated using a $(3+1)$ decomposition of spacetime, along with the corresponding 3-metric and extrinsic curvature. This formulation explicitly demonstrates that the BY mass depends on the choice of foliation or gauge.

It is important to highlight that, similar to the positive mass theorem, the positivity of this definition has been rigorously established in Ref.~\cite{Liu:2003bx}. The Brown-York framework effectively describes the total energy contained within a spherical region of radius $R$ surrounding a static gravitational source.

Additionally, it is worth noting that an alternative approach to defining gravitational field energy exists, known as the Lynden-Bell-Katz formalism~\cite{Bell:1985kz,Naresh:1986gg}. In this framework, the authors provide a definition for the gravitational field energy density specifically for the exterior Schwarzschild metric expressed in isotropic coordinates. One can demonstrates that the gravitational field energy calculated using both the Brown-York and Lynden-Bell-Katz prescriptions yields equivalent results for a Schwarzschild-de Sitter object when transforming from isotropic to curvature coordinates. What makes this particularly noteworthy is that these two distinct methodologies produce identical energy measurements, despite the well-known challenges in defining gravitational field energy unambiguously due to the absence of a covariant formulation. This observation suggests that for static spherically symmetric spacetimes, gravitational field energy can indeed be computed in a physically meaningful and intuitive way, with both approaches converging to the same result. Nevertheless, throughout our current investigation, we consistently employ the Brown-York formalism.

\subsection{Schwarzschild-de Sitter spacetime}
\label{s2.3}

It is important to note that the definition of metric in the static scenario represents the field of a stationary object, regardless of whether it is a black hole or any another kind of compact object. Therefore, the prescription of the metric is equally applicable to any static compact object or a Buchdahl star. Therefore we are now considering the Schwarzschild-de Sitter solution where the metric functions read as follows.
\begin{align}
    f = N = \sqrt{1-\frac{2M}{R}-\frac{\Lambda}{3}R^2} \,,
    \label{4}
\end{align}
where $\Lambda$ being the cosmological constant. We note that the horizons are located at the roots of polynomial under the square-root, this means at the horizon we have $f=0$. Note that we assume that there are two distinct horizons, the black hole horizon $R_{\rm BH}$ and the cosmological horizon $R_{\rm cos}$, which means that we assume $9M^2\Lambda < 1$. This is the discriminant which ensures that the polynomial has 3 real roots, two of which are located at $R>0$, these are the only physically meaningful ones.

For this case, one immediately finds that the MSH mass becomes
\begin{align}
    M_{\rm MSH} := \frac{R}{2}\big(1-f^2\big) =
    \frac{R}{2} \Big(\frac{2M}{R}+\frac{\Lambda}{3}R^2\Bigr) =
    \Big(M + \frac{\Lambda}{6}R^3\Big) \,,
    \label{5}
\end{align}
while mass definition~\eqref{2a} yields the simple result $\widetilde{M}=M$. This agrees with our short discussion regarding the inclusion or exclusion of the cosmological constant into mass definitions.

Likewise, the BY quasilocal mass is given by
\begin{align}
    M_{\rm BY} := R\Bigl(1-\sqrt{1-\frac{2M}{R}-\frac{\Lambda}{3}R^2}\Bigr) =
    R -\sqrt{R^2-2MR-\frac{\Lambda}{3}R^4} \,.
    \label{6}
\end{align} 
Let us remark that Brown-York mass cannot be easily adjusted to remove the effect of the cosmological constant and any such attempt would appear to be unnatural. Let us note that Eq.~\eqref{6} is only well defined when $R_{\rm BH} < R < R_{\rm cos}$. This condition does not impose any restrictions on our work since we are concerned with massive, static and spherically symmetric objects whose surface is located between the two horizons. From an astrophysical point of view, it is well-motivated to only consider such objects. For objects where the physical radius satisfies $R<R_{\rm BH}$ one would model a black hole. On the other hand, objects with $R>R_{\rm cos}$ would be of cosmological scales and thus not relevant in the present context.

For an asymptotic observer who remains indifferent to the cosmological constant, both the MSH and BY masses reduce to the bare ADM mass $M$ i.e., 
\begin{align}
    \lim_{R\to\infty} M_{\rm MSH}\Bigr|_{\Lambda=0} = M \,, 
    \quad \text{and} \quad 
    \lim_{R\to\infty} M_{\rm BY}\Bigr|_{\Lambda=0} = M \,.
    \label{7}
\end{align}
At the same time, the limiting case of $R\to\infty$ for the BY mass $M_{\rm BY}$ cannot be taken easily when $\Lambda > 0$ because of the cosmological horizon. Mathematically speaking, the argument of the square root (arising in~\eqref{6}) becomes negative when $R>R_{\rm cos}$.

On the other hand, one can also consider the limiting case as one approaches the horizon or the central singularity. As with the cosmological horizon, one should not evaluate the mass terms when $R<R_{\rm BH}$. Note that, formally, the final part of Eq.~\eqref{5} is well defined for all $R$. However, the definition~\eqref{2} is based on the metric function $f$, which is only well defined for $R_{\rm BH} < R < R_{\rm cos}$. Therefore, it is only meaningful to consider the limit $R \to R_{\rm BH}$ of the two definitions of the masses, which gives
\begin{align}
    \lim_{R \to R_{\rm BH}} M_{\rm MSH} = \frac{R}{2}\,, \qquad
    \lim_{R \to R_{\rm BH}} M_{\rm BY} = R \,.
\end{align}
Regardless of the mass definition, one can introduce the gravitational potential or surface potential in this context. In Eq.~\eqref{4}, we have the time-time component of the metric tensor as
\begin{align}
    -g_{00} = 1-\frac{2M}{R}-\frac{\Lambda}{3}R^2 =: 1 - 2\Phi\,.
    \label{9}
\end{align}
Therefore, the surface potential becomes
\begin{align}
    \Phi = \frac{M}{R}+\frac{\Lambda}{6}R^2 \,,
    \label{potential}
\end{align}
It is interesting to see that the choice of $\Phi$ naturally relates to the MSH mass Eq.~\eqref{5} in the sense that $\Phi = M_{\rm MSH}/R$, in analogy to Newtonian gravity. On the other hand, if we were to work with $\widetilde{M}$, one would have to add the cosmological constant back into the definition of the potential to achieve agreement with the metric function.

\section{Buchdahl bounds}
\label{s3}

The Buchdahl bound for a perfect fluid interior of an spherical object with mass $M$ and radius $R$ states that the potential $\Phi(R)$ felt by a radially infalling particle, $\Phi(R)=\frac{M}{R}\leq\frac{4}{9}$, whether the object is charged or neutral~\cite{Buchdahl:1959zz,Buchdahl:1966star}. This inequality is derived under very general conditions, where both interior matter density and isotropic pressure are positive, and the density decreases outward i.e., $\frac{d\rho}{dR}\leq0$. At the boundary where pressure vanishes, the interior metric is matched to the exterior Schwarzschild background. This limit also accounts for central pressure $P_c$ being theoretically finite i.e., $P_c<\infty$. This configuration corresponds uniquely to Schwarzschild’s interior solution as a solution of Einstein's equations. Numerous alternative derivations of the bound have since been proposed under various conditions and assumptions focusing on spherical and static compact stars within the framework of the Schwarzschild exterior vacuum spacetime~\cite{Sachs:1962wk,Bondi:1964zz,Islam:1969aa,Andreasson:2007ck,Karageorgis:2007cy,Giuliani:2007zza,Mak:2001ie,Cardoso:2019rvt}. The Buchdahl bound establishes a fundamental limit on the compactness of a stellar object, derived under broad and general assumptions. While certain specialised configurations may permit more compact matter distributions under specific constraints, recent rigorous studies have demonstrated that the Buchdahl bound cannot be violated for a fluid satisfying the energy conditions, maintaining radial stability, and retaining a subluminal sound speed~\cite{Alho:2022bki}. Consequently, any physically realistic matter distribution within a compact star must adhere to the Buchdahl bound. A Buchdahl star represents the extreme case where this bound is saturated, making it the most compact possible object without forming an event horizon. The bound can be further tightened by considering the dominant energy condition along with the requirement that the speed of sound remains subluminal~\cite{Barraco:2002ds,Fujisawa:2015nda}. Hence, the Buchdahl limit represents a fundamental constraint established under broad general conditions, although more compact configurations can occur under particular circumstances and specific conditions.
    
Recently a remarkable finding by one of us established that the Buchdahl bound can, in general, be expressed in terms of the gravitational field energy and non-gravitational matter energy for any static, charged, or neutral object~\cite{Dadhich:2019jyf}. This formulation relies entirely on the exterior geometry and does not consider the interior distribution of matter, regardless of its nature. Moreover, this approach suggests that the equilibrium within the Buchdahl star is governed by a relation similar to the gravitational Virial theorem, where the average kinetic energy equals half the average potential energy~\cite{Dadhich:2022kom}.

In what follows, the following two subsections will extend our examination of the Buchdahl compactness limit in the presence of a cosmological constant. This analysis will be conducted from two clearly defined and distinct viewpoints concerning Buchdahl bounds: the first approach considers the surface potential, while the second relies on the notion of gravitational field energy. Notably, this treatment will proceed without any dependence on interior matter configuration.

\subsection{Bounds using the gravitational potential}
\label{s3.1}

%

One can define a black hole region of an asymptotically flat spacetime to be the set of all events that cannot send causal signals to future null infinity. The event horizon of the black hole is the boundary of this region, it is a null hypersurface, generated by null geodesics that never reach future null infinity. In contrast, a Buchdahl star possesses a timelike boundary surface, characterised by its timelike normal vector. Therefore, to address a fixed value for the boundary radius, we instead rely on a straightforward physical property: the escape velocity of an infalling particle reaches unity at the black hole horizon. This condition is equivalent to the timelike Killing vector becoming null. On the other hand, the boundary radius of a Buchdahl star is determined by the condition $V^2_{\rm escape}=2\Phi(R)=8/9$. Therefore regarding the gravitational potential, the Buchdahl compactness limit is expressed as $\Phi(R)\leq\frac{4}{9}$, which holds true regardless of whether the object is charged or neutral~\cite{Buchdahl:1959zz}. This occurs because the reduction in the effective active gravitational mass i.e., $M_{\rm MSH}$ is exactly offset by a corresponding decrease in the object's radius $R$. The compactness constraint $\Phi(R)\leq\frac{4}{9}$ also implies a limit on the surface redshift being $Z_{\rm surface}\leq2$ and the escape velocity being $V_{\rm escape}\leq\sqrt{\frac{8}{9}}$. Therefore this suggests that any object, as long as it is not a black hole, will always exhibit a surface redshift $Z_{\rm surface}\leq2$, making this a definitive observational prediction.

In the standard Schwarzschild setting, the Buchdahl inequality can be derived directly from the constraint $\Phi(R) \leq 4/9$. Now it is always reasonable to assume that this bound on static object holds true in the presence of a cosmological constant term in the Schwarzschild setting. Therefore under this assumption, from Eq.~\eqref{potential} one may have
\begin{align}
    \frac{2M}{R}+\frac{\Lambda}{3}R^2 &\leq \frac{8}{9}\,\\
    \text{i.e.,}\, \quad M+\frac{\Lambda}{6}R^3 - \frac{4}{9}R &\leq 0 \,.
    \label{pot1}
\end{align}
Since we are interested to investigate the bounds in the presence of a cosmological constant term, we can write the above equation in a form that contains the metric function of the Schwarzschild-de Sitter solution. To do this, we begin by multiplying Eq.~\eqref{pot1} with $(-2/R)$ to arrive at
\begin{align}
    \sqrt{1-\frac{2M}{R}-\frac{\Lambda}{3}R^2} &\geq \frac{1}{3} \,.
\end{align}
When setting $\Lambda=0$, we find the bound in the Schwarzschild background, as stated by Buchdahl~\cite{Buchdahl:1959zz,Buchdahl:1966star}.    
                
\subsection{Bounds using the gravitational field energy and the Virial theorem}
\label{s3.2}

Let us briefly recall the prescription of the Brown-York quasilocal energy as mentioned in the previous subsection Sec.~\ref{s2.2}. The Brown-York quasilocal energy serves as our starting point due to its well-established significance. It is notable not only because it provides the anticipated result for the gravitational field energy in the first approximation, but also due to the significant attention it has received in the literature~\cite{Brown:2000dz,Blau:2007wj,Yu:2008ij}. Additionally, the positivity of the Brown-York quasilocal mass has been mathematically demonstrated by Liu {\it{et al.}} in Ref.~\cite{Liu:2003bx}.

Based on this formulation, it can be inferred that for a spherically symmetric static object, the bare ADM mass $M$ is infinitely distributed at spatial infinity. As the object undergoes gravitational collapse, it gathers not only the field energies resulting from gravitational effects but also contributions from the cosmological constant term.

The gravitational field energy is determined by taking the difference between the total mass-energy (represented by the quasilocal Brown-York mass-energy $M_{\rm BY}(r\leq R)$) and the non-gravitational matter energy (equivalent to the effective gravitational mass-energy $M_{\rm MSH}(r\leq R)$) enclosed within a radius $R$ of the object.

Thus, the absolute value of the gravitational field energy $E_{\rm G}$ being compared with the energy of positive matter, defined outside the radius $R$ is expressed as the difference between the two aforementioned masses (as detailed in subsections Sec.~\ref{s2.1} and Sec.~\ref{s2.2}). Hence, using Eqs.~\eqref{5} and~\eqref{6} we can define
\begin{align}
    E_G &:= M_{\rm BY} - M_{\rm MSH} 
    \nonumber \\ &=
    R - \sqrt{R^2 - 2MR -\frac{\Lambda}{3}R^4} - 
    \Bigl(M+\frac{\Lambda}{6}R^3\Bigr) \,.
    \label{Eg}
\end{align}
This definition is analogous to the one used in the Ref.~\cite{Dadhich:2019jyf}. It is remarkable to mention that this would imply that the compactness limit can be derived as a constraint on $E_{\rm G}$, which can be calculated solely using the exterior metric, independent of the interior structure of the object. 

On the other hand, by employing the BY quasilocal energy formulation to determine $E_{\rm G}$, previously it has been established that a black hole horizon can be characterised through the equipartition of total energy into the gravitational field energy (referred to as $E_{\rm G}$) and non-gravitational matter energy (referred as $E_{\rm P}$) components~\cite{Dadhich:1997ze}. Specifically, the horizon occurs where these two energy forms balance, i.e., $E_{\rm G} = E_{\rm P}$. This equipartition criterion provides a general and insightful definition of a black hole horizons based on energy considerations (for more details on this interpretation, see Refs.~\cite{Dadhich:1997ze},~\cite{Naresh:2015en}). Now in the context of a non-black hole compact objects, this naturally raises the question: Can we identify an analogous relation to determine the compactness limit for a Buchdahl star? The answer is the following: The gravitational energy $E_{\rm G}$ within such objects manifests as internal energy that opposes the gravitational effects of $E_{\rm P}$, with equilibrium achieved when these countervailing influences balance. Therefore it turns out that the Buchdahl compactness limit and the Buchdahl star solution emerge when the ratio $\gamma = E_{\rm G}/E_{\rm P}$ reaches a critical value, while $\gamma = 1$ for black holes. It is also noteworthy to mention that the results by Alho {\it{et al.}} in Ref.~\cite{Alho:2022bki} suggest that $\gamma$ varies continuously for elastic configurations and can approach $\gamma \to1$. However, for physically reasonable elastic objects $\gamma$ is close to the Buchdahl bound. 

The Virial theorem immediately suggests itself as a potential guide, as it governs equilibrium conditions through the ratio of kinetic to potential energy--specifically yielding a ratio of $1/2$ for systems of gravitationally interacting particles. In this framework, the internal energy (equivalent to $E_{\rm G}$) corresponds to kinetic energy, while $E_{\rm P}$ plays the role of potential energy. Remarkably, as it demonstrates, $\gamma = 1/2$ indeed reproduces the saturated Buchdahl limit $\Phi(R) = 4/9$ that defines the Buchdahl star~\cite{Dadhich:2019jyf}. This reveals that the equilibrium condition for a Buchdahl star mirrors the Virial relation. Thus, just as the black hole horizon marks where gravitational and non-gravitational energies equalise, the Buchdahl star represents where gravitational energy reaches half of non-gravitational energy--an equally profound and physically meaningful characterisation.

This demonstrates the fact that compactness limits can be determined purely from external spacetime properties. In what follows, now we illustrate how the compactness criterion of a Buchdahl star, characterised by the condition $\Phi(R) = 4/9$ necessarily implies the relation $E_{\rm G} = \frac{1}{2}E_{\rm P}$ in the presence of a cosmological constant. Likewise, it makes sense to identify the potential energy with 
\begin{align}
    E_{\rm P} := M_{\rm MSH}\,. 
\end{align}
Let us briefly comment on the fact that the gravitational field energy and the chosen potential energy could be defined differently, possibly yielding (slightly) different results. However, given the consistent results derived using the different approaches, one could consider our choice \emph{a posteriori} well motivated. 

Therefore, using Eqs.~\eqref{Eg} and~\eqref{5}, we have
\begin{align}
    \sqrt{R^2 - 2MR -\frac{\Lambda}{3}R^4} &=
    R - \frac{3}{2}\Bigl(M+\frac{\Lambda}{6}R^3\Bigr) \,.
\end{align}
In order to arrive at a bound similar to that derived when using the gravitational potential, the above mentioned equation needs to be factorised into an appropriate terms. This means we square the equation, collect terms and simplify to arrive at the following.
\begin{align}
    R\Bigl(M+\frac{\Lambda}{6}R^3\Bigr) &= \frac{9}{4}\Bigl(M+\frac{\Lambda}{6}R^3\Bigr)^2 \,.
\end{align}
We now have a common factor on both sides and therefore complete the above equation yields
\begin{align}
    -\frac{9}{4}\Bigl(M+\frac{\Lambda}{6}R^3\Bigr)
    \Bigl(M-\frac{4}{9}R+\frac{\Lambda}{6}R^3\Bigr) &= 0 \,.
    \label{Virialfinal}
\end{align}

As previously discussed, the energy within an object's interior increases by an amount corresponding to its gravitational energy. This increment serves as a measure of its internal energy, which may take the form of kinetic or binding energy. In the case of a fluid distribution, this internal energy can manifest as pressure. Alternatively, for specific scenarios such as Vlasov kinetic matter, it may appear as thermal (kinetic) energy. Moreover, it can also manifest in various other forms depending on the physical context. Therefore the equilibrium within a stellar interior is determined by the balance between gravitational energy (internal/kinetic) and non-gravitational matter (potential) energy. Hence the Buchdahl bound condition implies that the internal energy must not exceed half of the non-gravitational energy. A Buchdahl star is defined when this condition is saturated, meaning its equilibrium state is characterised by internal energy being exactly half of the non-gravitational (matter) energy. Therefore from Eq.~\eqref{Virialfinal} for the Buchdahl bound condition $E_{\rm G} \leq \frac{1}{2} E_{\rm P}$, we can write
\begin{align}
  -\frac{9}{4}\Bigl(M+\frac{\Lambda}{6}R^3\Bigr)
  \Bigl(M+\frac{\Lambda}{6}R^3-\frac{4}{9}R\Bigr) &\geq 0 \,.
  \label{Virialfinal2}
\end{align}
Let us now use the expression for the gravitational potential $\Phi$ introduced in Eq.~\eqref{potential}. This yields
\begin{align}
  -\frac{9}{4}\Phi R \Bigl(\Phi R-\frac{4}{9}R\Bigr) =
  -\frac{9}{4} R^2 \Phi \Bigl(\Phi -\frac{4}{9}\Bigr) 
  \geq 0 \,.
  \label{Virialfinal2a}
\end{align}
which for a positive cosmological constant yields the Buchdahl analogue compactness bound in the presence of a cosmological constant
\begin{align}
    \Phi - \frac{4}{9} \leq 0 \,,
    \label{Virialfinal3} \\
    \text{i.e.}\quad \frac{M}{R}\leq\frac{4}{9}-\frac{\Lambda}{6}R^2 \,.
    \label{Virialfinal4} 
\end{align}
Therefore, the inequality obtained from analysing the gravitational field energy and applying the Virial theorem aligns with the result presented in Eq.~\eqref{pot1}, which was derived solely from considerations of the gravitational potential. This establishes a direct correspondence of Buchdahl bounds with the Virial equilibrium condition, even in the presence of a cosmological constant $\Lambda$.

Additionally it is important to mention that in the limit of maximum compactness, it is plausible that the matter interaction within the fluid may begin to weaken, leading to the formation of pure kinetic Vlasov matter--a system of freely moving particles interacting solely through gravity. In such a scenario, the internal energy would indeed correspond to kinetic energy, and the equilibrium would then be dictated by the well-known Virial theorem. Interestingly, for this Vlasov kinetic distribution, the Buchdahl bound can only be saturated in the case of a thin shell that satisfies the strong energy condition~\cite{Andreasson:2007ck}.

It is also crucial to emphasise that compactness should inherently depend on the internal structure, which includes the equation of state of the fluid distribution and its binding energy. Essentially, the properties of the fluid itself dictate the degree of compactness an object can attain, and this principle underlies the derivation of the compactness bounds. Such an inquiry is indeed highly intuitive. So far, we have explored the Buchdahl compactness limit in the context of a cosmological constant, which is expressed entirely in terms of the gravitational potential and the gravitational field energy. These quantities are uniquely determined by the exterior metric. Moving forward, we will now examine the generalised Buchdahl bound derived from the Schwarzschild interior solution, incorporating a cosmological constant and assuming constant energy density. Subsequently, we will extend our analysis in the following sections.
 
\subsection{Bound using the Schwarzschild interior solution}
\label{s4}

The Buchdahl inequality for the Schwarzschild interior solution, assuming constant density, in the presence of a cosmological constant $\Lambda$ was first derived in~\cite{Boehmer:2003uz}. Using the usual assumption of finite central pressure, $P_c < \infty$, analogous to the standard case without cosmological term, the following inequality was derived
\begin{align}
    \sqrt{1 - \frac{2M}{R} - \frac{\Lambda}{3} R^2} \geq 
    \frac{1}{3} - \frac{\Lambda}{12 \pi \rho} \,, 
    \label{buch-lambda}
\end{align}
with subsequent developments and extensions found in Refs.~\cite{Balaguera-Antolinez:2004ytv,Boehmer:2005sm,Balaguera-Antolinez:2005ngl}. It is immediately clear that the implied mass-radius bounds become weaker as the value of $\Lambda$ increases. For the special value $\Lambda=4\pi\rho$ the right-hand side of~\eqref{buch-lambda} vanishes and the condition no longer implies the non-existence of horizons. 

Since the cosmological constant can be viewed as the energy density of the dark energy fluid, one concludes that the right-hand side of Eq.~\eqref{buch-lambda} is indeed dimensionless. In order to extract an equality that only involve mass $M$, radius $R$ and the cosmological constant $\Lambda$, we use $\rho=3M/(4\pi R^3)$, then we can write
\begin{align}
  \sqrt{1-\frac{2M}{R}-\frac{\Lambda}{3}R^2} \geq 
  \frac{1}{3} - \frac{R^3\Lambda}{9M} \,.
\end{align}
After simplifying the above equation, we arrive at
\begin{align}
    1-\frac{2M}{R}-\frac{\Lambda}{3}R^2 &\geq
    \frac{1}{9} - \frac{2}{3}\frac{R^3\Lambda}{9M} + 
    \frac{R^6\Lambda^2}{81M^2} \,,
    \label{ineq10}
\end{align}
which can be written as a product of three terms as follows
\begin{align}
  -2\left(M+\frac{\Lambda}{6}R^3 \right)
  \left[M-\frac{2R}{9}\left(1-\sqrt{1-\frac{3\Lambda }{4}R^2}\right)\right]
  \left[M-\frac{2R}{9}\left(1+\sqrt{1-\frac{3\Lambda }{4}R^2}\right)\right] \geq 0 \,.
  \label{factor1}
\end{align}
This is not obvious and was first reported in~\cite{Boehmer:2005sm} where this factorisation was used to determine a minimum mass bound depending on the cosmological constant. However, one can easily check that $M=-\Lambda R^3/6$ is a root of the cubic Eq.~\eqref{ineq10}, which can then be reduced to a quadratic equation. The roots of that quadratic are given by the terms in the square brackets. One should also note that to satisfy this inequality mentioned in~\eqref{factor1}, the 3 factors can either have signs $(+,+,-)$ or $(-,-,-)$, so that one requires an even number of minus or plus signs. Given that $(M + \Lambda R^3/6) > 0$ for a positive cosmological constant term (or a small negative cosmological constant), we conclude that the two terms in the square brackets have different signs. 

To extract a Buchdahl type inequality, let us now consider the final term of Eq.~\eqref{factor1} by assuming $\Lambda \ll 1$. One finds
\begin{align}
  M-\frac{2R}{9}\left(1+\sqrt{1-\frac{3\Lambda }{4}R^2}\right) &\approx
  M-\frac{4}{9}R + \frac{\Lambda}{12}R^3 \leq 0 \,,
  \label{expansion1}
\end{align}
which is equivalent to 
\begin{align}
  \frac{M}{R} \leq \frac{4}{9} - \frac{\Lambda}{12}R^2 \,.
  \label{expansion2}
\end{align}
It is intriguing to note that the $\Lambda$ term differs by a factor of one half from the corresponding result in Eq.~\eqref{Virialfinal4}.

Eqs.~\eqref{pot1} and~\eqref{Virialfinal2} both include the term $\Lambda R^3/6$, while the interior solution produces a factor of $\Lambda R^3/12$. Since the formulation of the gravitational field energy with the Virial theorem and the bound on the gravitational potential are in agreement -- both are based on strong physical arguments -- this highlights a non-trivial tension with the interior solution. 

When studying mass-radius bounds for charged objects~\cite{Dadhich:2019jyf}, one also finds different inequalities depending on the approach taken when the cosmological constant is included. This implies that these different approaches depend on different implicit assumptions, and it is not clear where precisely these differences originate. Interestingly, it is not clear whether a sharper inequality derived using the exterior geometry, for example, compared to the inequality derived in the interior, should have implications on the allowed matter configurations. It might be possible, however, to extract other inequalities from the constant density interior solution by changing the boundary conditions that were used. One possible idea would be to set $\bar{\rho}=\rho_0 + \Lambda/(8\pi)$ and $\bar{P}=P-\Lambda/(8\pi)$ and to use the condition $\bar{P}(R=0)=P_c$ as the definition of the central pressure. Likewise, one can ask the same question about the appropriate definition of the boundary of the object, let this be $\bar{P}(R)=0$ which would mean $P(R)=-\Lambda/(8\pi)$ instead of $P(R)=0$. These small changes will affect the compactness bounds differently, and it would be interesting to understand whether these bounds will agree with the inequality derived using the exterior metric only.

\section{Discussions}

The primary objective of our investigation is to highlight a novel and insightful approach for determining an accurate generalised Buchdahl-analogue compactness limit for static relativistic compact objects in the presence of a cosmological constant. This limit incorporates gravitational field energy and is derived exclusively from the exterior solution, without any dependence on the interior configuration. It is well-known that the Buchdahl compactness bound for any static relativistic astrophysical object necessitates the gravitational field energy to remain below half of its non-gravitational matter energy, thereby upholding the Virial theorem. Intriguingly, the equality between gravitational field energy and non-gravitational matter energy characterises the black hole horizon. As discussed in Ref.~\cite{Dadhich:1997ze}, timelike particles undergo gravitational acceleration driven by non-gravitational matter energy, whereas photons experience only spatial curvature induced by gravitational field energy~\cite{Naresh:2015en}. As a timelike particle approaches the horizon, its velocity tends to the speed of light, effectively behaving like a photon. Consequently, at the horizon, the contributions from both sources—non-gravitational matter energy (responsible for acceleration) and gravitational field energy (responsible for curvature)—must balance. This equipartition principle defines the horizon's location and thereby motivates our exploration of a generalised compactness limit of a static horizon-less compact object in the presence of a non-zero cosmological constant, employing the Virial theorem and the framework of gravitational field energy. 

Very recently, a most insightful balance of energy inside and outside the object has been envisaged by one of us~\cite{Dadhich:2025sjr}. At infinity, a system is infinitely dispersed having the bare ADM mass $M$. It then picks up gravitational energy, $E_{\rm out}=E_{G}$ as it collapses, and which lies exterior to the object, while the energy inside is $E_{\rm in} = M - E_{G}$. Now there are three possible equilibrium configurations: (a) At infinity, $E{\rm out}=0$; this means $E_{\rm in} =M$, (b) $E_{\rm in}=0$; i.e., $M=E_{G}$, and (c) $E_{\rm in}=E_{\rm out}$; which is equivalent to $2M=E_{G}$. The case (b), which is dual to the case (a), defines the black hole horizon while case (c) characterises the Buchdahl star. That is, the relativistic Virial theorem prescribes that in general relativity there are only two possible equilibrium states of limiting compactness one, a black hole with a null boundary or a Buchdahl star with timelike boundary. It is remarkable that, in contrast to the Newtonian case, the GR Virial theorem prescribes compactness, one for null and the other for the timelike particles. It is most remarkable that such a simple general consideration leads to the profound prediction that gravitational collapse can either end in a Buchdahl star or a black hole.

Furthermore, at the boundary of such a compact object, the binding energy matches the gravitational field energy exterior to the object. These quantities can be viewed as complementary, with their equivalence at the boundary serving as a constraint that determines the compactness limit. This reveals a profound connection between the interior and exterior dynamics, implying that an object's compactness is restricted only by the condition that its gravitational or binding energy does not surpass half of its non-gravitational matter energy. Remarkably, our analysis computes this limit using solely the static exterior Schwarzschild-de Sitter metric, whereas binding energy depends on the interior fluid distribution and the equation of state, leading to varied solutions based on fluid properties.

It is natural to ask whether the $\Lambda$-dependent Buchdahl bounds have implications for the trapping polytropes studied in~\cite{Stuchlik:2016xiq,Stuchlik:2017qiz,Stuchlik:2025fws}. These configurations can contain stable null circular geodesics in their interior even when the surface radius is far outside the exterior photon sphere. The trapping property is not directly controlled by the global compactness $M/R$, nor by the surface Buchdahl condition. The Buchdahl inequality therefore does not give a criterion for the existence of trapping zones. Its relevance is instead indirect: it gives a global exterior compactness constraint on admissible configurations, while $\Lambda$ also affects whether the polytropic configuration exists at all. This is an interesting interplay where $\Lambda$ affects the existence of two distinct properties of solutions which should be studied further.

As a future extension of this work, it would be compelling to investigate rotating compact objects, which are more realistic and significant from an astrophysical point of view. However, rotation introduces considerable mathematical challenges. Recent progress in this direction has been made in~\cite{Hernandez-Pastora:2017fmg,Herrera:2017tkq} where an interior solution that smoothly matches to the Kerr solution has been reported. It would be interesting to apply the methods of our current work to the rotating case. For slowly rotating interiors, solutions exist under approximation schemes~\cite{Hartle:1968si}, and extensive studies~\cite{Meinel:2008kpy} demonstrate numerical solutions for both rotating fluids and kinematic matter sources. Recent numerical simulations have also explored dynamically stable ergostars~\cite{Tsokaros:2019mlz}. Additionally, exploring compactness limits for generalised elastic stars in the presence of a cosmological constant, possibly derived from the exterior geometry, presents another intriguing avenue for future research which has not been explored yet.

\subsection*{Acknowledgement}

CGB expresses gratitude to USTC for their warm hospitality. CGB and ND would like to thank the University of KwaZulu Natal for their kind hospitality, where this project was initiated. SD acknowledges the financial support received through the USTC Fellowship Level A--CAS-ANSO Scholarship 2024 (formerly the ANSO Scholarship for Young Talents) for doctoral studies. Furthermore, SD extends sincere thanks to Yi-Fu Cai for coordinating seminar at the Particle Cosmology group (COSPA), USTC during the course of this work.

\addcontentsline{toc}{section}{References}
\bibliographystyle{jhepmodstyle}
\bibliography{main}

\end{document}